\documentclass[useAMS,usenatbib]{mn2e}
\usepackage[dvips]{graphicx,color}
\usepackage{txfonts}
\usepackage{amssymb}
\usepackage{indentfirst}
\usepackage{epsfig}

\newcommand{\Ka}{\ensuremath{\hbox{K}\alpha~}}

\newcommand{\A}{\AA~}

\def\cha{{\it Chandra }}

\begin{document}
\title{The intrinsic line width of the Fe \Ka line of AGN} 

\author[J. Liu et al.]{Jiren Liu\thanks{E-mail: jirenliu@nao.cas.cn}  \\
	 $^{}$National Astronomical Observatories, 20A Datun Road, Beijing 100012, China\\
}

\date{}

\maketitle

\begin{abstract}

X-ray fluorescent lines are unique features of the reflection spectrum
of the cold torus when irradiated by the central AGN. 
Their intrinsic line widths can be used to probe the line-emitting region.
The line widths of the Fe \Ka line measured from the first order \cha 
High Energy Grating (HEG) spectra are $3-5$ times larger than those measured with the
Si \Ka line for Circinus, Mrk 3, and NGC 1068.
Because the observed Si \Ka and Fe \Ka lines are not
necessarily coming from the same physical region, it is uncertain whether
the line widths of the Fe \Ka line are over-estimated or not.
We measured the intrinsic line widths of the Fe \Ka line of several nearby bright AGN 
using the second and 
third order \cha HEG spectra, whose spectral resolutions are better than the first order data.
We found the measured widths are all smaller than those from the first
order data. The results clearly show that the widths 
of the Fe \Ka line measured from the first order HEG data are over-estimated.
It indicates that the Fe \Ka lines of the studied sources are 
originating from regions around the cold dusty torus.

\end{abstract}


\section{Introduction}

The cold dusty gas around active galactic nuclei (AGN) obscures and reprocesses the intrinsic
radiation of AGN and is the key ingredient of understanding different types of AGN
\citep[e.g.][]{Ant93}. The obscuring gas emits fluorescent lines when irradiated 
by the central AGN. The most prominent one is the Fe \Ka line at 6.4 keV, which is 
found to be ubiquitous in all types of AGN \citep[e.g.][]{NP94,Shu10,Fuk11}.
As a result, the fluorescent Fe \Ka line can be used to probe the properties
of the obscuring gas. For example, the stability of the Fe \Ka line of most AGN
indicates that the obscuring gas should be outside the broad line region (BLR) \citep[e.g.][]{Bia12}. 
The exact emitting region of the fluorescent line 
can be inferred from the intrinsic width of the Fe \Ka line \citep[e.g.][and reference therein]{Shu11}. 

The measurement of the intrinsic line width, however, is limited by the spectral 
resolution of currently available instruments.
The first order data of \cha High Energy Transmission 
Grating Spectrometer \citep[HETGS,][]{Can05} provide a spectral resolution of 0.012 \A 
(full width half maximum, FWHM) with its High Energy Grating (HEG), which
at 6.4 keV corresponds to $\sim1860$ km\ s$^{-1}$, very close to the measured mean FWHM
of the Fe \Ka line (2000 km\ s$^{-1}$) by \citet{Shu11}. 

Because the HEG spectral resolution is higher at lower energies, the measurement 
of the line width can be improved by using other low-energy fluorescent lines.
In a recent paper \citep{Liu16b}, we measured the FWHM of the Si \Ka line (at 1.74 keV)
for Circinus, Mrk 3, and NGC 1068, which
are $570\pm240$, $730\pm320$, and $320\pm280$ km\ s$^{-1}$, respectively.
These values are $3-5$ times smaller than those measured from the Fe \Ka line previously,
and the estimated line-emitting regions are outside the dust sublimation radii of these AGN.
It indicates that the intrinsic widths of the Fe \Ka line are likely to be over-estimated.
Nevertheless, because the Si \Ka line is much more sensitive to absorption than the Fe \Ka line,
the Si \Ka line may not come from the same physical region as the Fe \Ka line.
This makes the estimation of the emitting region of the Fe \Ka line still uncertain.

In principle, the uncertainty of the Fe \Ka emitting region can be solved by 
spectroscopic observations with resolution higher than the first order HEG data. We noted
that the second and third order \cha HEG data have a resolution approximately 2 and 3 times better 
than the first order data, but with a less effective area \citep{Can05}. 
We checked the second and third order \cha HEG data of Circinus, Mrk 3, 
and NGC 1068, and found that indeed, the higher order HEG data provide more stringent constraints
on the intrinsic widths of the Fe \Ka line than the first order data. 
In this letter we present these measurements.
Besides Circinus, Mrk 3, and NGC 1068, we also include NGC 3783, NGC 4151, NGC 4388, and NGC 4151, 
the higher order HEG data of which are deep enough to provide a 
meaningful measurement of the intrinsic width of the Fe \Ka line.
The errors quoted are for 90\% confidence level.

\begin{table}
	\caption{The line widths of the Fe \Ka line measured from the second and third order 
	\cha HEG data and the first order data}
\begin{tabular}{lccccccccc}
\hline
\hline
Name&$\sigma$($\pm2$,$\pm3$)&$\sigma$($\pm1$)&$v_{\rm FWHM}$($\pm2$,$\pm3$)&$v_{\rm FWHM}$(Si K$\alpha$)$^a$\\
			 & eV &   eV  & km\ s$^{-1}$&km\ s$^{-1}$\\
\hline
Circinus & $6.2\pm1.4$  & $9.8\pm0.9$ & $680\pm150$ & $570\pm240$\\
NGC 4151 & $11.4\pm6.9$ & $18.6\pm2.9$& $1260\pm760$& $610\pm240$\\
NGC 3783 & $11.3\pm5.2$ & $14.9\pm3.0$& $1250\pm570$& -\\
Mrk 3    & $4.9^{+5.2}_{-4.9}$  &$19.0\pm3.9$& $540^{+570}_{-540}$& $730\pm320$\\
NGC 1068 & $7.1^{+7.1}_{-7.1}$  &$18.0\pm4.1$& $780^{+780}_{-780}$& $320\pm280$\\
NGC 4388 & $10.0^{+7.6}_{-10.0}$  &$13.0\pm7.5$& $1100^{+840}_{-1100}$& -\\
NGC 4507 & $11.9^{+21.5}_{-11.9}$ &$14.6\pm7.0$& $1310^{+2370}_{-1310}$& -\\
\hline
\end{tabular}
\begin{description}
\begin{footnotesize}
\item
  Note: For Circinus, NGC 4151, and NGC 3783, the second and third order \cha HEG data are deep enough 
  to allow the $\chi^2$ statistic, 
  while for other sources, the Cash statistic is used.
  The $\chi^2$ statistic is used for the first order data of all sources. 
  $^a$The FWHM of the Si \Ka line is quoted from \citet{Liu16b}, except for that of NGC 4151, 
  which is measured in this paper, while for NGC 3783, NGC 4388 and NGC 4507, the Si \Ka line is 
  too weak to provide meaningful constraints.
\end{footnotesize}
\end{description}
\end{table}

\section{Observational data }

Because the \cha HEG effective areas of the second and third order are about 15 times less than 
that of the first order \citep{Can05}, only for nearby bright sources with deep exposures, 
the higher order HEG data can be used to measure the intrinsic width of the Fe \Ka line.
We searched \cha Transmission Grating Data Archive and Catalog
\citep[TGCat,][]{TG}, and 
found that besides Circinus, Mrk 3, and NGC 1068, the second and third order HEG data of 
NGC 3783, NGC 4151, NGC 4388, and NGC 4151 are also usable.
Among them, NGC 3783 and NGC 4151 are classified as type 1.5 AGN, while the others are type 2 AGN.
All the spectra are extracted from a region with a 2 arcsec half-width in the cross-dispersion
direction. The instrumental responses are extracted using TGCat software with the 
calibration database (CALDB) 4.6.8.
The HEG data from $\pm2$ and $\pm3$ orders are combined together.
The continua around the Fe \Ka line of NGC 4151 of different observations 
show variations with a factor of 5, 
and only the low-state observations are used. The continuum variations of NGC 3783
between different observations are within a factor
of 2, and to improve the signal-to-noise (S/N) ratio, all the observations of NGC 3783 are used.
For all the other sources, no apparent continuum variations are noted.
Some higher order HEG spectra of stellar-mass black holes have been published in \citet{Mil15,Mil16}.

\begin{figure}
\includegraphics[width=2.85in]{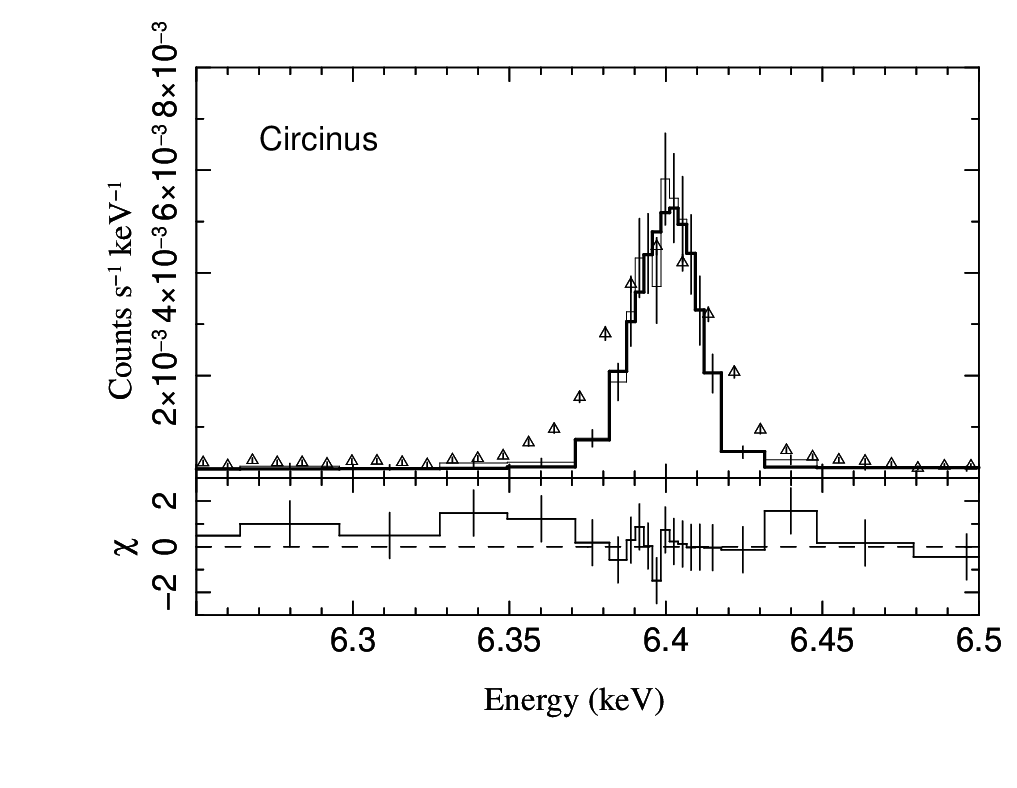}
\includegraphics[width=2.85in]{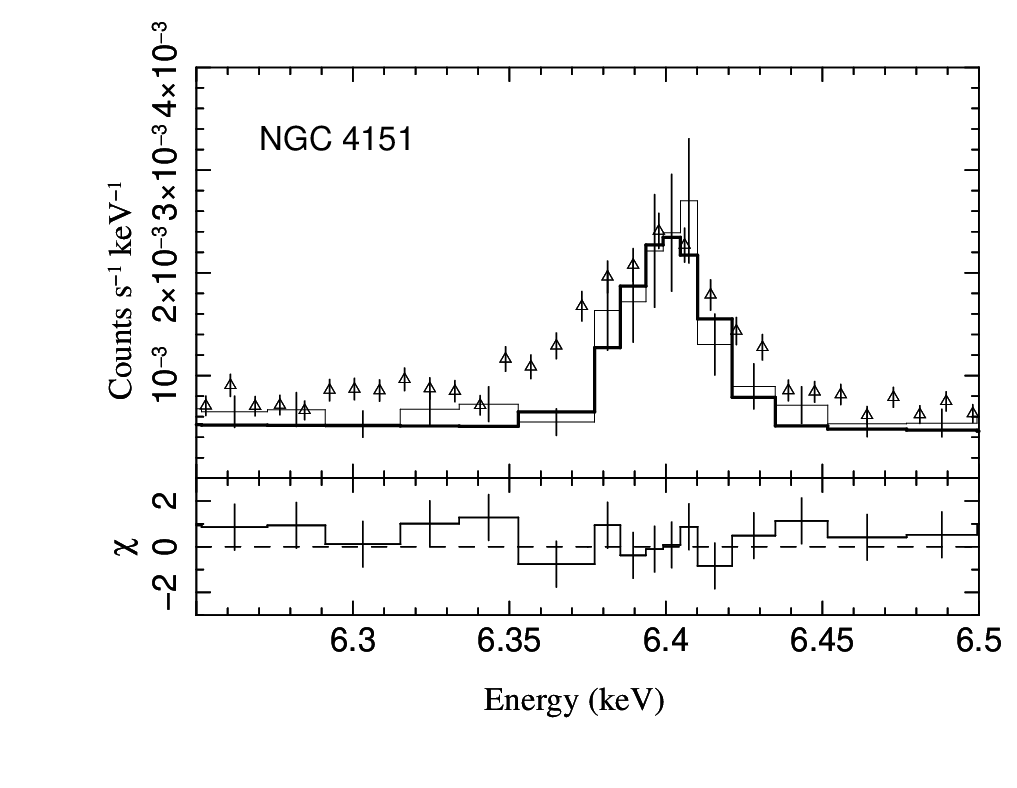}
\includegraphics[width=2.85in]{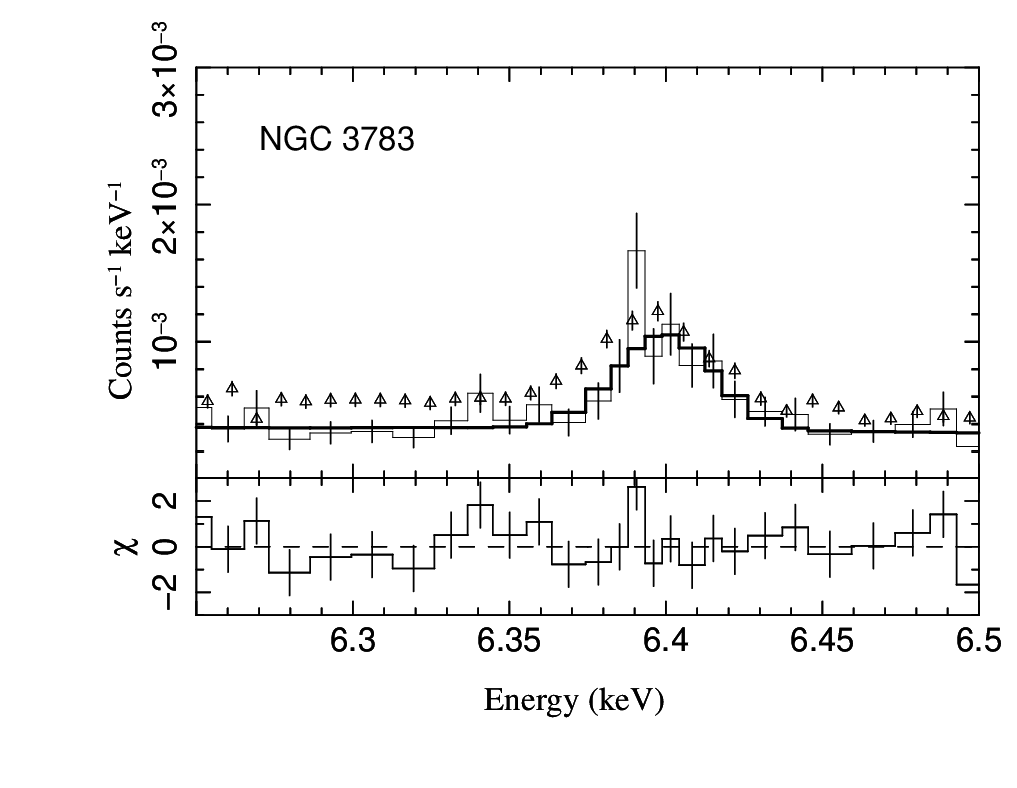}
\caption{
The second and third order \cha HEG spectra of the Fe \Ka line 
of Circinus, NGC 4151, and NGC 3783, corrected for redshift.
The data are rebinned with a minimum S/N of 4.
The fitted models of two Gaussian lines plus a linear continuum 
are plotted as the thick solid histograms. For comparison, the first order HEG 
spectra (reduced by a factor of 15) are over-plotted as triangles. 
}
\end{figure}

\begin{figure*}
\includegraphics[width=2.85in]{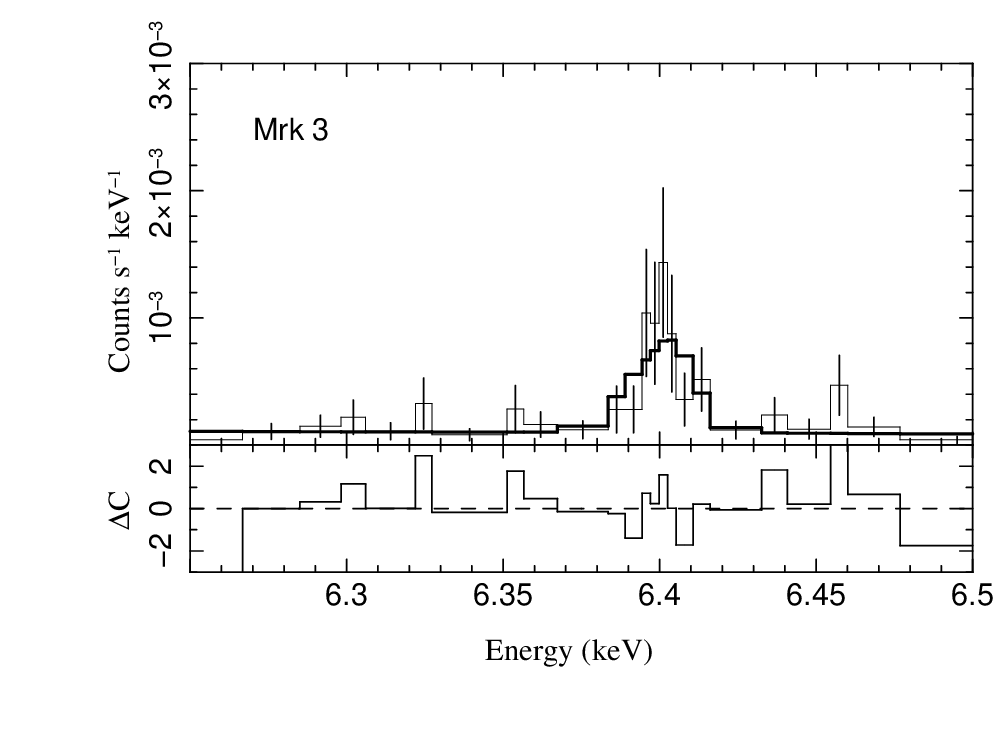}
\includegraphics[width=2.85in]{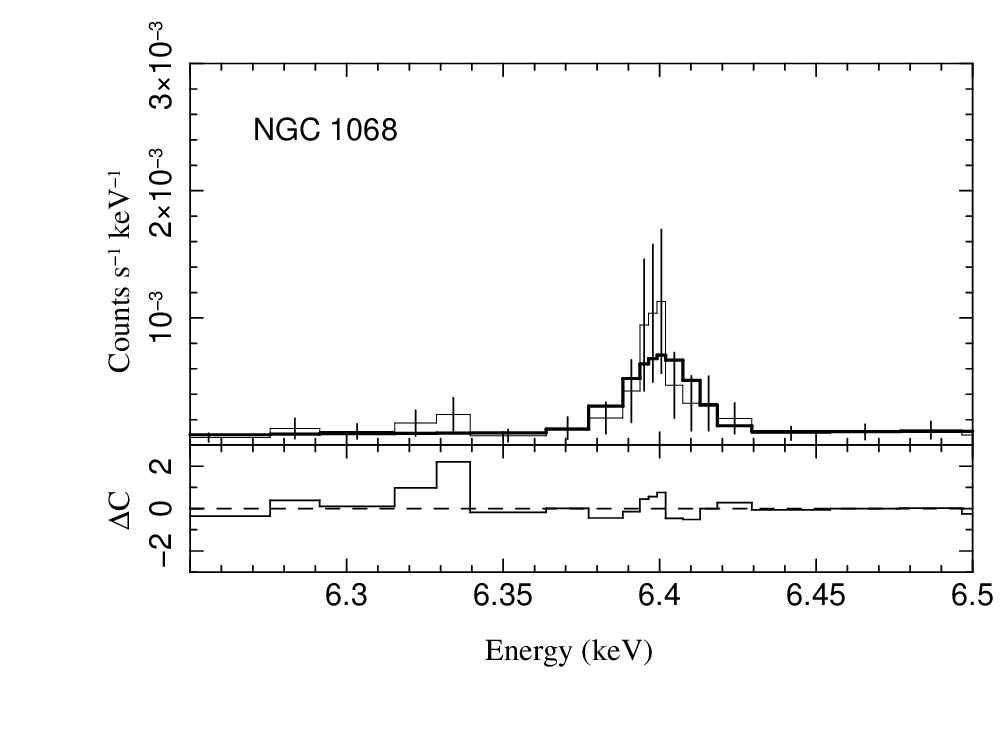}
\includegraphics[width=2.85in]{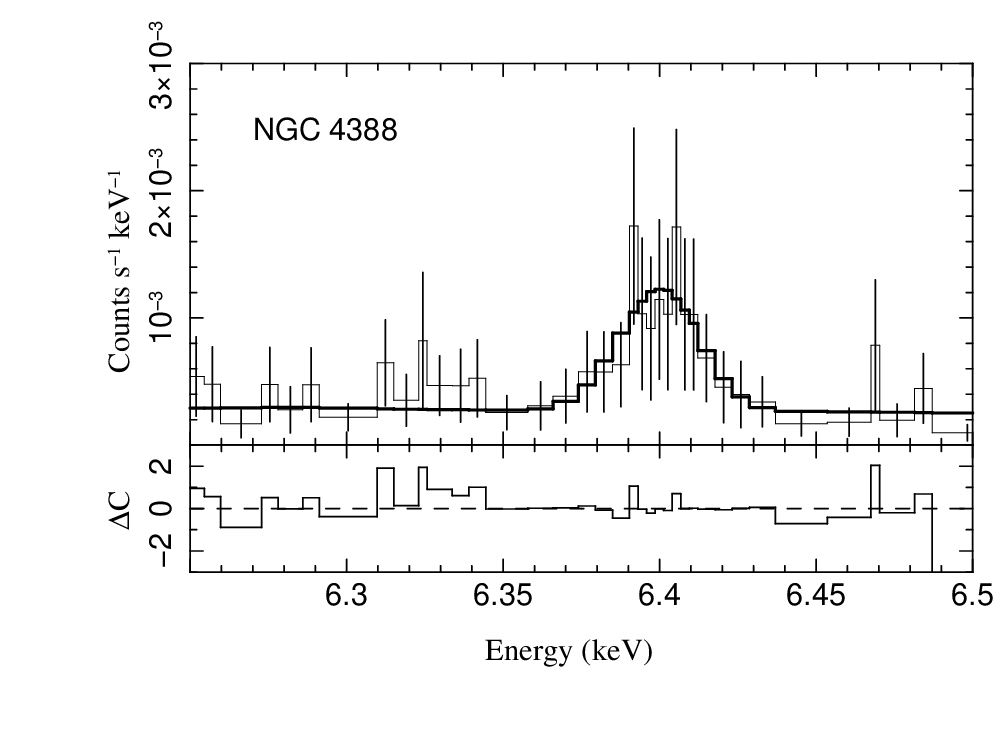}
\includegraphics[width=2.85in]{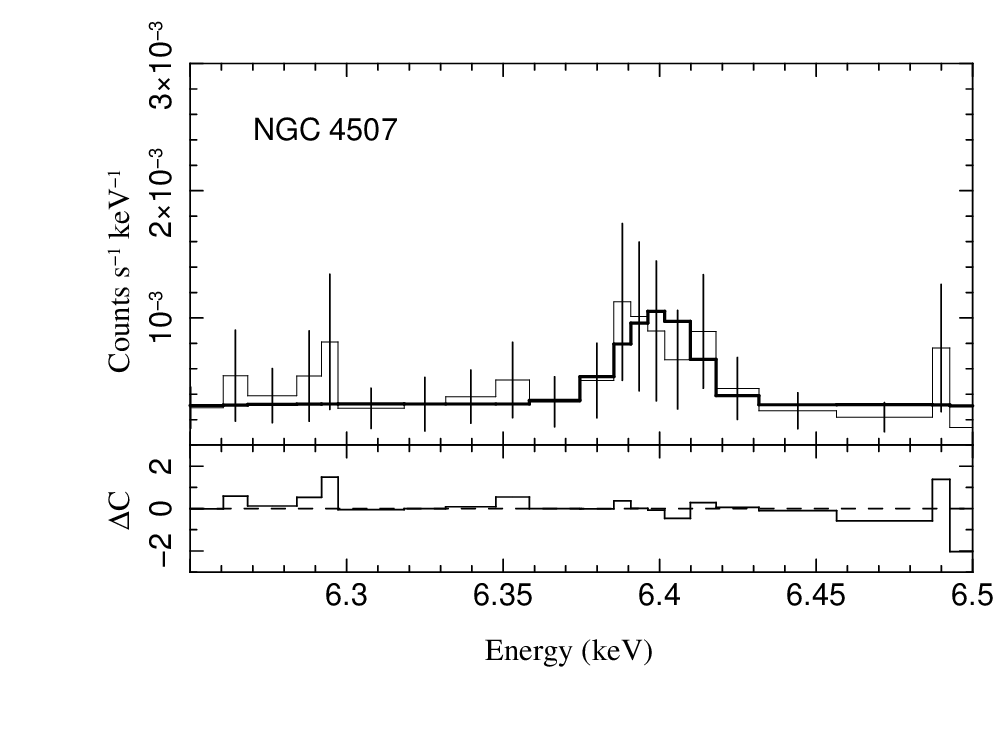}
\caption{
The second and third order \cha HEG spectra of the Fe \Ka line 
of Mrk 3, NGC 1068, NGC 4388, and NGC 4507, corrected for redshift.
The data are rebinned with a minimum S/N of 1.5.
The fitted models of two Gaussian lines plus a linear continuum 
are plotted as the thick solid histograms. 
}
\end{figure*}

\section{Results}

For Circinus, NGC 4151, and NGC 3783, the second and third order \cha HEG spectra are
deep enough to allow a robust measurement of the width of the Fe \Ka line. Their spectra 
are plotted in Figure 1,  which are rebinned with a minimum S/N ratio
of 4. For comparison, the corresponding first order HEG spectra (reduced by a factor of 15)
are over-plotted in Figure 1. As can be seen, the second and third order HEG spectra 
are narrower than the first order spectra, and can provide better constraints on the
width of the Fe \Ka line. It also shows that the Fe \Ka line is dominated by the neutral
Fe atoms, and the contribution from low-ionized Fe$^+$ ions 
\citep[located around 6.42 keV,][]{KM93} is negligible.

The neutral Fe \Ka line is composed of a doublet, K$\alpha_1$ at 6.404 keV,
and K$\alpha_2$ at 6.391 keV, with a flux ratio of 2:1 \citep{Bea67}.
We model the observed spectra with two Gaussian lines plus a linear continuum.
The two Gaussian lines are centered at 6.404 and 6.391 keV with their redshifts and
widths fixed with each other, and the intensity of the K$\alpha_2$ line
is set to be half of the K$\alpha_1$ line.
The fitting region is between 6 and 6.6 keV.
The fitted results are plotted in Figure 1 and listed in Table 1.
For comparison, the line widths obtained from the first order HEG data using the same
model are also listed in Table 1. Note that the widths of the first order HEG data presented here
are a little smaller than those reported by \citet{Shu10,Shu11}, because they used one Gaussian
line to model the Fe \Ka doublet. For Circinus, the one with the best data, 
the line width measured from the second and third order spectra is about 2/3 
of that from the first order spectra. The corresponding FWHM velocity is similar to that 
obtained from the Si \Ka line in \citet{Liu16b}. For NGC 4151 and NGC 3783, the line widths of the 
second and third order spectra are also about 2/3 of those from the first order spectra.

The second and third order \cha HEG spectra of all the other sources are plotted in Figure 2.
Because their signals are not as good as the above three sources, their spectra
are rebinned with a minimum S/N ratio of 1.5, and the Cash statistic is used. The same doublet model
is adopted. The fitted results are plotted in Figure 2 and listed in Table 1.
As expected, the errors of the measured line widths are larger, compared with 
those of Circinus, NGC 4151, and NGC 3783. Nevertheless, all the line widths of the
second and third order spectra are smaller than those of the first order spectra.
The corresponding FWHM velocities of the Fe \Ka line of Mrk 3 and NGC 1068 are consistent with
those obtained from the Si \Ka line in \citet{Liu16b}.

\section{Conclusion and Discussion}

X-ray fluorescent lines are unique features of the reflection spectrum 
emitted by the torus when irradiated by the central AGN. 
The intrinsic line widths of the X-ray fluorescent 
lines can be used to probe the location of the line-emitting region. 
The line widths of the Fe \Ka line measured from the first order \cha HEG
spectra are $3-5$ times larger than those measured with the 
Si \Ka line for Circinus, Mrk 3, and NGC 1068.
Nevertheless, because the observed Si \Ka and Fe \Ka lines are not 
necessarily coming from the same physical region, it is still uncertain whether 
the line widths of the Fe \Ka line are over-estimated or not.

In this letter we measured the line width of the Fe \Ka line using the second and 
third order \cha HEG spectra
of several nearby bright AGN. For Circinus, NGC 4151, and NGC 3783, which
have deep enough signals, the measured line widths are about 2/3 of those measured 
from the first order HEG spectra.
While for Mrk 3, NGC 1068, NGC 4388, and NGC 4507, the measured line widths are also smaller than
those measured from the first order HEG spectra.
The results clearly show that the line widths of the Fe \Ka line measured from the first order
HEG spectra are over-estimated.

In principle, the first order and higher order HEG data should provide consistent results 
on the line widths. However, when the intrinsic line width is too
narrow to be resolved by the first order data, we do not expect
the same results, since there are observational noises and the instrument responses are not ideal.
To validate our measurements, we simulated 100 HEG observations of the Fe \Ka line using 
Marx\footnote{http://space.mit.edu/cxc/marx} \citep{Dav12} for the two Gaussian model
of Circinus with an intrinsic width of 5 eV (corresponding to the velocity of that measured
from the Si \Ka line width), and the exposure time is set to be the same as that of the real
observations.
We note that Marx is used to generate the \cha HEG RMF calibration products, and 
contains all known effects about the instrumental line profile (D. Huenemoerder, private
communication). 
We then fitted the simulated HEG spectra with the same two Gaussian model.
We found that 90\% of the fitted Fe \Ka widths from the first order spectra are within 
$8.2\pm1.1$ eV. On the other hand, 90\% of the fitted widths from the 
second and third order spectra are within $6.1\pm1.4$ eV.
Considering that the first order spectra are contaminated by the Compton shoulder, which is less
important for the higher order spectra of better resolution, the simulation results
are fully consistent with our measured results.

We note that except for Circinus, NGC 4151, and NGC 3783, the Fe \Ka lines of all the 
other sources are not well resolved, and the measurements presented here are only upper 
limits of the true widths. Deeper data with similar or higher spectral resolution are needed 
to obtain the true widths.

For the current measurements, the FWHM velocities of the Fe \Ka line are around 
1000 km/s, similar to those of narrow emission lines.
If assuming a virialized orbit, 
the Fe \Ka line-emitting regions are close to the dust sublimation radii,
as inferred from the widths of Si \Ka line in \citet{Liu16b}.
It indicates that the Fe \Ka lines of the studied sources are 
originating from regions around the cold dusty torus.

\section*{Acknowledgements}
We thank our referee for valuable comments that improved the work
and David Huenemoerder for assistance on the higher order 
HEG calibrations. 
JL is supported by NSFC grant 11203032.
This research is based on data obtained from the \cha Data Archive.

\bibliographystyle{mn2e}

\end{document}